\documentclass[12pt]{iopart}
\usepackage{mathrsfs}
\expandafter\let\csname equation*\endcsname\relax
\expandafter\let\csname endequation*\endcsname\relax
\usepackage{amsmath}
\usepackage{amsfonts}
\usepackage{amssymb}
\usepackage{amsthm}
\usepackage{graphicx}
\usepackage{color}
\usepackage{hyperref}
\usepackage{bm}
\usepackage[caption=false]{subfig}
\usepackage{verbatim}

\definecolor{Red}{rgb}{1,0,0}

\newcommand{\Imag}{{\Im\mathrm{m}}}   

\newcommand{\ve}{\boldsymbol} 

\begin{document}
\title[The magnetic field driven superconductor--metal transition ...]{The magnetic field driven superconductor--metal transition in disordered hole-overdoped cuprates}

\author{Lina G. Johnsen}
\address{Center for Quantum Spintronics, Department of Physics, \\Norwegian University of Science and Technology, NO-7491 Trondheim, Norway}

\begin{abstract}
By solving the Bogoliubov--de Gennes equations for a $d$-wave superconductor, we explore how the interplay between disorder and the orbital depairing of an external magnetic field influences the superconductor--metal transition of the hole-overdoped cuprates. For highly disordered systems, we find granular Cooper paring to persist above the critical field where the superfluid stiffness goes to zero. We also show that because the vortices are attracted to regions where the superconducting pairing is already weak, the Caroli--de Gennes--Matricon zero-bias peak in the local density of states at the vortex cores disappears already at moderate disorder.
\end{abstract} 


\maketitle

\section{Introduction}

The rich phase diagram of the cuprates allow for the study of a variety of different phases by adjusting the concentration of dopants \cite{damascelli_rmp_03,alloul_rmp_09}. While undoped cuprates are antiferromagnetic Mott insulators, a sufficient level of doping can cause the onset of high-temperature superconductivity \cite{bednorz_zpb_86}.
In the hole-underdoped and optimally doped regimes, superconductivity can no longer be described by Bardeen-Cooper-Schrieffer (BCS) theory \cite{bardeen_pr_57}. In these regimes, the normal state is not a Fermi-liquid, but rather a pseudogap phase or a strange metal \cite{alloul_rmp_09,varma_prl_89,alloul_prl_89,rossat_pc_91}. As the temperature is decreased, the onset of superconductivity is determined by the superfluid stiffness rather than the Cooper pairing \cite{emery_nat_95}. Moreover, the competition between superconducting and antiferromagnetic order causes magnetic structures to arise around impurities \cite{julien_prl_00,wang_prl_02,andersen_prl_07} and vortex cores \cite{lake_sci_01,demler_prl_01,zhu_prl_01,lake_nat_02,hoffman_sc_02,zhu_prl_02}.

In the less studied hole-overdoped regime, a large Fermi surface with well defined quasi-particles makes Fermi-liquid theory a more suitable description of the normal-state \cite{plate_prl_05,vignolle_nat_08}. 
In the superconducting state, experimental observations fit better with BCS theory \cite{alloul_rmp_09,simard_prb_19}.
However, some puzzling observations include Cooper pairs forming above the superconducting critical temperature ($T_c$) in Bi$_2$Sr$_2$CaCu$_2$O$_{8+x}$ (Bi2212) \cite{gomes_nat_07,pasupathy_sci_08,he_prx_21}, and a large fraction of uncondenced electrons below $T_c$ in several of the hole-overdoped cuprates \cite{uemura_nature_93,niedermayer_prl_93,rourke_nat_11,bozovic_nat_16,mahmood_prl_19}.
Recent experimental studies of La$_{2-x}$Sr$_x$CuO$_4$ (LSCO) \cite{bozovic_nat_16,wu_nat_17,bozovic_jsnm_18,herrera_prb_21} found that the superfluid stiffness decreases linearly with increasing temperature, and that the critical temperature depends on the zero-temperature superfluid stiffness. The claim that these findings go beyond BCS theory has been challenged by theoretical works \cite{lee-hone_prb_17,lee-hone_prr_20,wang_prl_22}.
No consensus has yet been reached, but it is clear that the relation between the superconducting pairing and superfluid stiffness shows interesting properties even when described within the dirty BCS framework.

As a prime example, Li \textit{et al.} \cite{li_npj_21} predicted that upon increasing the concentration of non-magnetic impurities in hole-overdoped Bi2212, the superfluid stiffness is lost, and the superconductor transitions into a state with granular Cooper pairing and spontaneous supercurrent loops.
Since the hole-overdoped cuprates are $d$-wave superconductors, they are not protected by Anderson's theorem \cite{anderson_jpcs_59,abrikosov_jetp_61,balatsky_rmp_06} and superconductivity is strongly suppressed in the vicinity of non-magnetic impurities \cite{franz_prb_97,ghosal_prb_00}. In Refs.~\cite{he_prx_21,li_npj_21} modelling Bi2212, the authors in particular highlight the importance of flat bands in the anti-nodal regions causing increased pair-breaking and the transition into the granular state.

In this work, we study how the interplay between disorder and the orbital depairing of an external magnetic field influences the superconductor--metal transition of the hole-overdoped cuprates. By solving the Bogoliubov--de Gennes (BdG) equations for a disordered $d$-wave superconductor, we study the superfluid stiffness and superconducting pairing close to the transition. Like Li \textit{et al.} \cite{li_npj_21}, we consider a band-structure fitting experimental measurements of Bi2212 \cite{he_prx_21} with a band-filling that gives rise to flat bands close to the Fermi level in the antinodal regions. 
We find that when the system becomes sufficiently disordered, granular Cooper pairing persists beyond the magnetic field driven superconductor to metal transition.
This allows us to conveniently reach the intermediate regime predicted for the disorder driven superconductor--metal transition in Ref.~\cite{li_npj_21} by tuning an external magnetic field.
Moreover, we show that the Caroli--de Gennes--Matricon (CdGM) zero-bias peak in the local density of states (LDOS) at the vortex cores vanishes at moderate disorder, as the vortices start penetrating regions where the superconductivity is already weak. This sensitivity to disorder could contribute to the elusiveness of the CdGM zero-bias peak in experimental studies \cite{renner_prl_98,hoogenboom_pcs_00,pan_prl_00,levy_prl_05,yoshizawa_jpsj_13}.

\section{Model}

We consider a disordered type-II $d$-wave superconductor under an applied magnetic field. The lattice structure considered is a two-dimensional (2D) square lattice, which to good approximation models the quasi-2D structure of the cuprates \cite{damascelli_rmp_03}. Moreover, we assume the superconducting film to be thin enough that the orbital effect of the perpendicular magnetic field dominates over the Zeeman splitting.
This system can be described by the Hamiltonian
\begin{align}
    H =& -\sum_{\ve{i},\ve{j},\sigma}t_{\ve{i},\ve{j}}e^{i\phi_{\ve{i},\ve{j}}}c_{\ve{i},\sigma}^{\dagger}c_{\ve{j},\sigma}-\sum_{\ve{i},\sigma}\left(\mu-V_{\ve{i}}\right)n_{\ve{i},\sigma}
    +\sum_{\left<\ve{i},\ve{j}\right>} \left(\Delta_{\ve{i},\ve{j}}c_{\ve{i},\uparrow}^{\dagger}c_{\ve{j},\downarrow}^{\dagger}+\text{h.c.}\right).
    \label{eq:Hamiltonian}
\end{align}
Here, $c_{\ve{i},\sigma}^{\dagger}$, $c_{\ve{i},\sigma}$, and $n_{\ve{i},\sigma}=c_{\ve{i},\sigma}^{\dagger}c_{\ve{i},\sigma}$ are the creation, annihilation, and number operators associated with a spin-$\sigma$ electron at lattice site $\ve{i}$. Each of the above terms are explained in the following.

The first term describes hopping between neighboring lattice sites. 
We include hopping between nearest, next nearest and third nearest neighbors. These three types of hopping are associated with hopping parameters $t_{\ve{i},\ve{j}}=t$, $t'$, $t''$, respectively. 
The applied magnetic field introduces an accumulated Peierls phase
\begin{align}
    \phi_{\ve{i},\ve{j}}=-\frac{\pi}{\Phi_{0}^{\text{SC}}}\int_{\ve{r}_{\ve{j}}}^{\ve{r}_{\ve{i}}}d\ve{r}\cdot \ve{A}(\ve{r})
\end{align}
when an electron moves from position $\ve{r}_{\ve{j}}$ to position $\ve{r}_{\ve{i}}$. Here, $\Phi_0^{\text{SC}} = hc/2e$ is the superconducting flux quantum, and $\ve{A}(\ve{r})=B(0,x,0)$ is the vector potential in the Landau gauge resulting from a homogeneous external magnetic field $B$.

The second term in Eq.~\eqref{eq:Hamiltonian} introduces the chemical potential $\mu$ and the disorder potential $V_{\ve{i}}$. We consider a random disorder potential in the range $V_{\ve{i}}\in[-V,V]$. The chemical potential is adjusted in order to fix the hole density $x$ while considering different disorder strengths. The hole density is given by 
\begin{align}
    x = \frac{1}{N_x N_y}\sum_{\ve{i},\sigma}\big<1-n_{\ve{i},\sigma}\big> .
\end{align}
We consider hole-doped superconductors ($0<x\leq1$) far away from half-filling ($x=0$). In this regime, the cuprates are purely superconducting without competing antiferromagnetic order. Experimental observations suggest a more conventional behavior, where BCS theory captures many aspects of the superconductivity well \cite{damascelli_rmp_03,alloul_rmp_09}.

The last term in Eq.~\eqref{eq:Hamiltonian} introduces the superconducting pairing arising from a nearest-neighbor interaction described within the mean field approximation \cite{zhu_book_16}. The pairing correlation $\Delta_{\ve{i},\ve{j}}=J\left<c_{\ve{i},\uparrow}c_{\ve{j},\downarrow}\right>$ is used to calculate the spin-singlet pairing $\Delta^{\text{S}}_{\ve{i},\ve{j}}=(\Delta_{\ve{i},\ve{j}}+\Delta_{\ve{j},\ve{i}})/2$. The $d$-wave spin-singlet paring is defined as
\begin{align}
    \Delta_{\ve{i}}^{\text{d}}&=\frac{1}{4}\big(\Delta_{\ve{i}}^{+\ve{x}}+\Delta_{\ve{i}}^{-\ve{x}}-\Delta_{\ve{i}}^{+\ve{y}}-\Delta_{\ve{i}}^{-\ve{y}}\big),
    \label{eq:dwave}
\end{align}
where $\Delta_{\ve{i}}^{\pm\ve{x}(\ve{y})}=\Delta_{\ve{i},\ve{i}\pm\ve{x}(\ve{y})}\exp\big(i\phi_{\ve{i},\ve{i}\pm\ve{x}(\ve{y})}\big)$. 
In order to make the numerical calculations feasible, we need to scale down the lattice size compared to a realistic system. The parameter $J$ should ideally be chosen large enough that the vortex diameter is much smaller than the width of the system, but still small enough that the vortex spans at least a few lattice sites. The parameters chosen for each plot is given in the corresponding figure text.
We consider the zero-temperature limit as our theoretical framework do not capture the effect of thermal fluctuations. 

We calculate the spin-singlet $d$-wave pairing self-consistently from the Bogoliubov-de Gennes equations of the Hamiltonian in Eq.~\eqref{eq:Hamiltonian}. We follow the method in Ref.~\cite{zhu_book_16}. For details, see the \hyperref[appendix]{Appendix}. In order to reduce the system size without disturbance from edge effects, we define a magnetic unit cell containing an even number of superconducting flux quanta and apply periodic boundary conditions at its edges. Thus, we can solve the BdG equations for a periodic array of $M_x \times M_y$ magnetic unit cells of size $N_x \times N_y$.

In highly disordered materials, the existence of superconducting pairing is no longer a good measure for whether the material is superconducting. This is because paring can exist locally without any global phase coherence. For defining the superconducting phase transition, we therefore introduce the superfluid stiffness $D_{\text{s}}$. We calculate the superfluid stiffness from the Kubo formula \cite{scalapino_prl_92,scalapino_prb_93,zhu_book_16}
\begin{align}
    \frac{D_{\text{s}}}{\pi e^2}=&\left<-K_x \right>-\Lambda_{xx}(q_x = 0, q_y \to 0, \omega=0),\label{eq:Kubo1}
\end{align}
that describes the linear response to a vector potential $A_x  e^{i(\ve{q}\cdot\ve{r}_{\ve{i}}-\omega t)}$ applied in the $x$ direction.
Above, $\left<K_x \right>$ is the expectation value of the kinetic energy and $\Lambda_{xx}(\ve{q},\omega)$ is the current-current correlation function. The kinetic energy associated with the $x$ oriented bonds is given by 
\begin{align}
    K_x =-\frac{1}{N_x N_y}\sum_{\ve{i},\ve{\delta},\sigma}\delta_x^2 \big(t_{\ve{i}+\ve{\delta},\ve{i}}e^{i\phi_{\ve{i}+\ve{\delta},\ve{i}}}c_{\ve{i}+\ve{\delta},\sigma}^{\dagger}c_{\ve{i},\sigma}+\text{h.c.}\big).
\end{align}
The current-current correlation function is given by 
\begin{align}
    \Lambda_{xx}(\ve{q},\omega)&=\frac{i}{N_x N_y}\int_0^{\infty}dt\:e^{i\omega t}\left<\left[J_x (\ve{q},t),J_x (-\ve{q},0)\right]\right>,
\end{align}
where $J_{x}(\ve{q},t)=\sum_{\ve{i}}\exp(-i\ve{q}\cdot\ve{r}_{\ve{i}})J_x(\ve{r}_{\ve{i}},t)$ is the Fourier transform of the $x$ oriented particle current
\begin{align}
    J(\ve{r}_{\ve{i}},t)=i\sum_{\ve{\delta},\sigma}\delta_x \big(t_{\ve{i}+\ve{\delta},\ve{i}}e^{i\phi_{\ve{i}+\ve{\delta},\ve{i}}}c_{\ve{i}+\ve{\delta},\sigma}^{\dagger}c_{\ve{i},\sigma}-\text{h.c.}\big).
\end{align}
With these physical quantities, we are able to study whether superconducting pairing is present, whether the material is superconducting, and how currents flow inside the material.
Finally, we define the local density of states in terms of the retarded Green's function \cite{zhu_book_16}
\begin{align}
    N_{\ve{i}} &= -\frac{1}{\pi}\sum_{\sigma}\Imag\left[G^{\text{R}}_{\ve{i},\sigma,\ve{i},\sigma}(\omega)\right],\\
    G^{\text{R}}_{\ve{i},\alpha,\ve{j},\beta}(\omega) &= -i\int_{0}^{\infty}dt\:e^{i\omega t}\big<\big\{c_{\ve{i},\alpha}(t),c_{\ve{j},\beta}^{\dagger}(0)\big\}\big>.
\end{align}
This furthermore allows us to study the local density of states inside the vortex cores.

\section{Results}

\begin{figure}[t]
    \centering
    \includegraphics[width=0.9\textwidth]{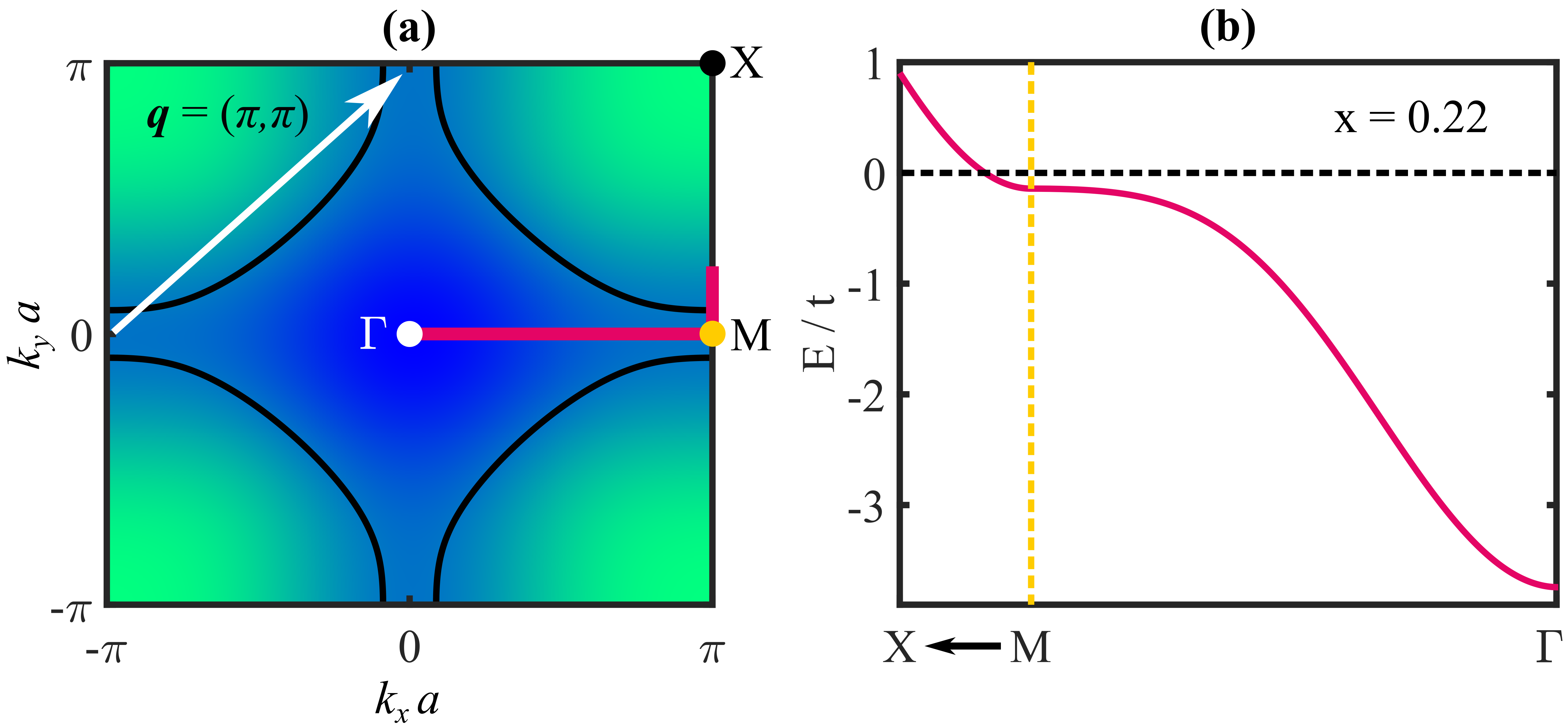}
    \caption{We consider a normal state band structure fitting experimental measurements of Bi2212 at 22 \% hole doping \cite{he_prx_21} using next nearest and third nearest neighbor hopping parameters $t'/t = -0.05$ and $t''/t = 0.2$, respectively \cite{li_npj_21}. Panel a) shows the Fermi surface (black lines), and panel b) the bandstructure along the red line in panel a). The band structure is plotted from the $\Gamma$ point to the M point and towards the X point. The position of the $\Gamma$, M, and X points in the first Brillouin zone are indicated by a white, yellow, and black dot, respectively. In panel b), the Fermi level is marked by the black dotted line, and the M point is marked by the yellow dotted line. Scattering between the antinodal regions by a wave vector $\boldsymbol{q}=(\pm\pi,\pm'\pi)$, as illustrated by the white arrow in panel a), is pair breaking due to the $d$-wave pairing having opposite signs in the antinodal regions around $(\pm\pi,0)$ and $(0,\pm\pi)$ \cite{he_sci_14}. The flat band shown in panel~b) increases the scattering between the antinodal regions, thus making the $d$-wave pairing more sensitive to impurities \cite{zou_arxiv_21}.}
    \label{fig:01}
\end{figure}

In order to study how the strong pair breaking associated with the $d$-wave pairing symmetry affects the magnetic field driven superconducting transition, we choose parameters modelling Bi2212 at 22\% hole doping. As shown in Fig.~\ref{fig:01}, this band filling gives rise to flat bands close to the Fermi surface in the antinodal regions, causing increased scattering between regions where the superconducting pairing has opposite signs \cite{he_prx_21}. For the case of zero external magnetic field, such flat bands have been shown to cause a strong suppression of the superconducting pairing and superfluid stiffness under increasing disorder \cite{li_npj_21}. By choosing parameters giving a high sensitivity to disorder, our parameters allow us to study the opposite limit compared to the robust conventional superconductor studied in Ref.~\cite{datta_arxiv_21}.

\begin{figure}[t]
    \centering
    \includegraphics[width=0.975\textwidth]{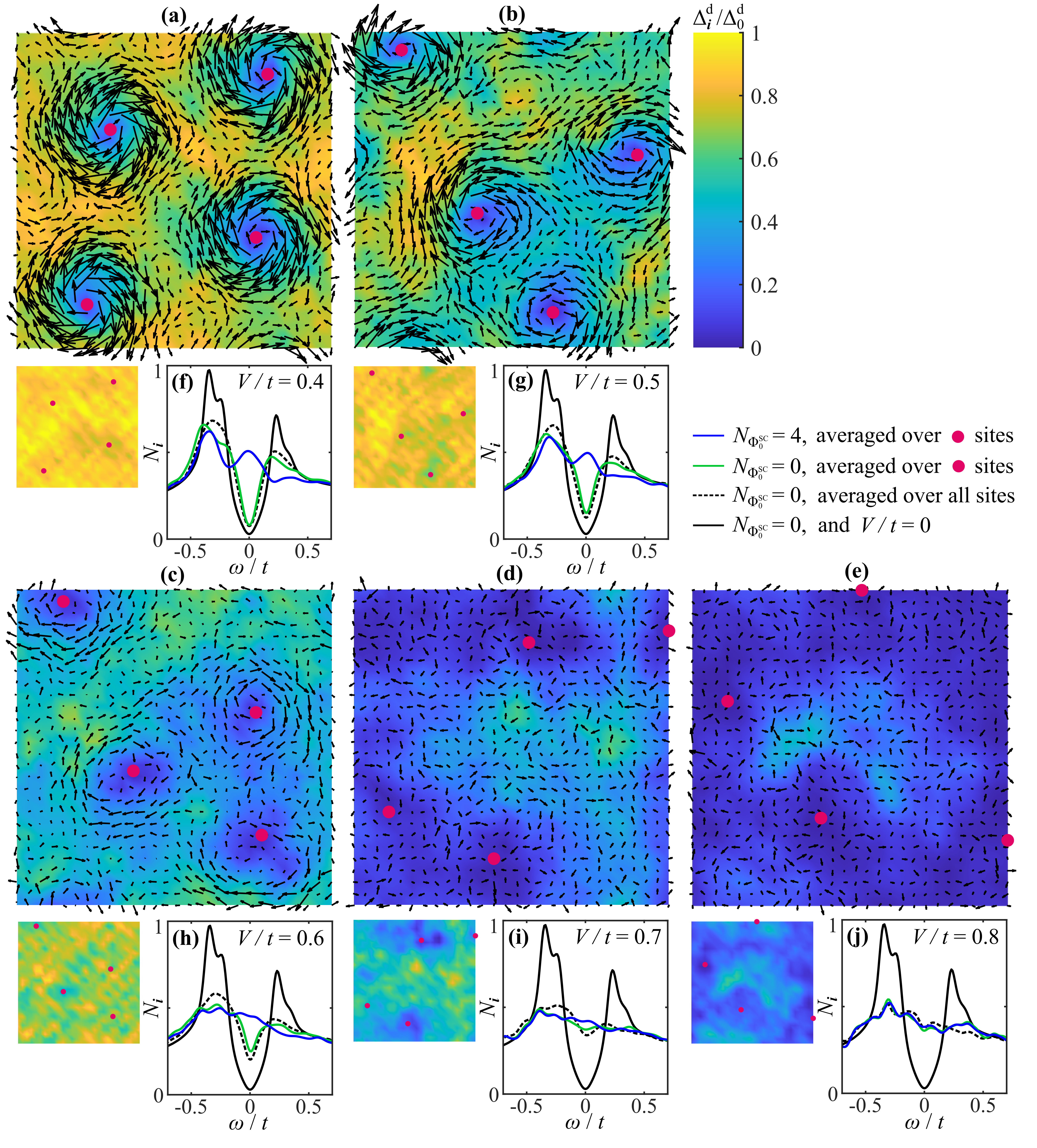}
    \caption{Panel (a)-(e): The $d$-wave pairing $\Delta_{\ve{i}}^{\text{d}}$ in a magnetic unit cell containing $4$ vortices for impurity potentials $V/t=0.4$, $0.5$, $0.6$, $0.7$, and $0.8$, respectively. The corresponding plots for zero magnetic field are shown below each panel. The pairing is scaled by its value $\Delta^{\text{d}}_0$ in a clean system without vortices. The arrows represent the net current through each lattice site, and the red dots mark the positions of the vortex cores determined from the phase of $\Delta_{\ve{i}}^{\text{d}}$. Panel (f)-(j): The local density of states at the vortex cores in panel (a)-(e) (blue), and at the corresponding lattice sites for zero magnetic field (green). The LDOS is averaged over all vortex sites and their nearest and next nearest neighbors. The black curves show the LDOS averaged over all lattice sites in the absence of the magnetic field for the given impurity potential (dotted) and in a clean system (solid). (Parameters: $J/t = 0.9$, $N_{x(y)} = 28$, $M_{x(y)} = 10$.)}
    \label{fig:02}
\end{figure}

We first consider the how the superconducting pairing evolves under increasing disorder in the presence of a constant magnetic field. Fig.~\ref{fig:02}(a)-(e) show the superconducting pairing inside a magnetic unit cell penetrated by four superconducting flux quanta for various disorder strengths. While the vortices in a clean system form a regular lattice due to the
mutual repulsion between vortices, increasing disorder causes the vortices to shift towards
highly disordered regions where the superconducting pairing is already weak. In Fig.~\ref{fig:02}(a)-(e), we typically find one vortex where the superconducting pairing is at its weakest, and the other three vortices in or close to local minima sufficiently far away from other vortices. In the highly disordered systems, a vortex can be located close to a grain boundary if the vortex repulsion makes this energetically favorable. It is however very unlikely to find vortices in the middle of a superconducting grain. The magnetic field therefore suppresses the Cooper pairing in the regions where the pairing is already weak, and causes superconducting pairing to survive in grains surrounded by regions where the pairing is absent. We will later show that this granularity is associated with a vanishing superfluid stiffness.

In Fig.~\ref{fig:02}(f)-(j), we study how the local density of states at the vortex cores changes as the disorder increases. When the disorder is low, the LDOS at the vortex core show a clear Caroli--de Gennes--Matricon zero-bias peak \cite{caroli_pl_64}. However, the CdGM zero-bias peak vanishes already at moderate disorder where the pairing is not yet granular in the absence of magnetic fields. This is clearly seen from Fig.~\ref{fig:02}(c) and (h). The zero-bias peak is nearly absent, although before applying the magnetic field there is a clear superconducting gap in the average density of states and the superconducting paring always remains above 40\% of its value in the clean system.
Since the vortices are being attracted to regions of high disorder where the superconducting pairing is minimal, the suppression of the zero-bias peak is determined by the disorder potential in the most strongly disordered regions. In these regions, we see that the superconducting gap in the LDOS in the absence of an external magnetic field is more filled up than when we average over the whole system, see especially panel~(h). 
The sensitivity to disorder could be a contributing factor to the absence of the CdGM zero-bias peak in hole-overdoped cuprates.
The zero-bias peak has been observed in conventional superconductors \cite{hess_prl_89} and more recently also in the cuprate YBa$_2$Cu$_3$O$_{7-\delta}$ (Y123) \cite{berthod_prl_17}. Observing the CdGM zero-bias peak in cuprates have otherwise proved difficult, and experimental studies of Bi2212 have not shown signatures of a robust zero-bias peak \cite{renner_prl_98,hoogenboom_pcs_00,pan_prl_00,levy_prl_05,yoshizawa_jpsj_13}.

In Fig.~3, we plot the superconducting pairing and superfluid stiffness as a function of the disorder strength and the applied magnetic field. Each data point is calculated by averaging over all lattice sites and $70-100$ impurity configurations. The error bars represent the standard deviation. The standard error of the mean is $12\%-10\%$ of the standard deviation. Since we are considering relatively small lattice sizes, the superconducting transition is sensitive not only to the magnetic field and the impurity strength, but also the impurity configuration. Although the superconducting transition is sharper for a specific impurity configuration, the superconducting transition is seen in our plots as a gradual transition where an increasing fraction of the impurity configurations result in zero superfluid stiffness. This is represented by the error bars dropping down to zero and the data points gradually approaching zero. Note that when plotting the superconducting pairing and superfluid stiffness as a function of the magnetic field, we will not get a purely monotonous decrease, particularly for weaker disorder strengths. This is because not all of the field strengths can produce a square lattice of vortices in a clean system. This error decreases with increasing system size and disorder.

\begin{figure}[b]
    \centering
    \includegraphics[width=\textwidth]{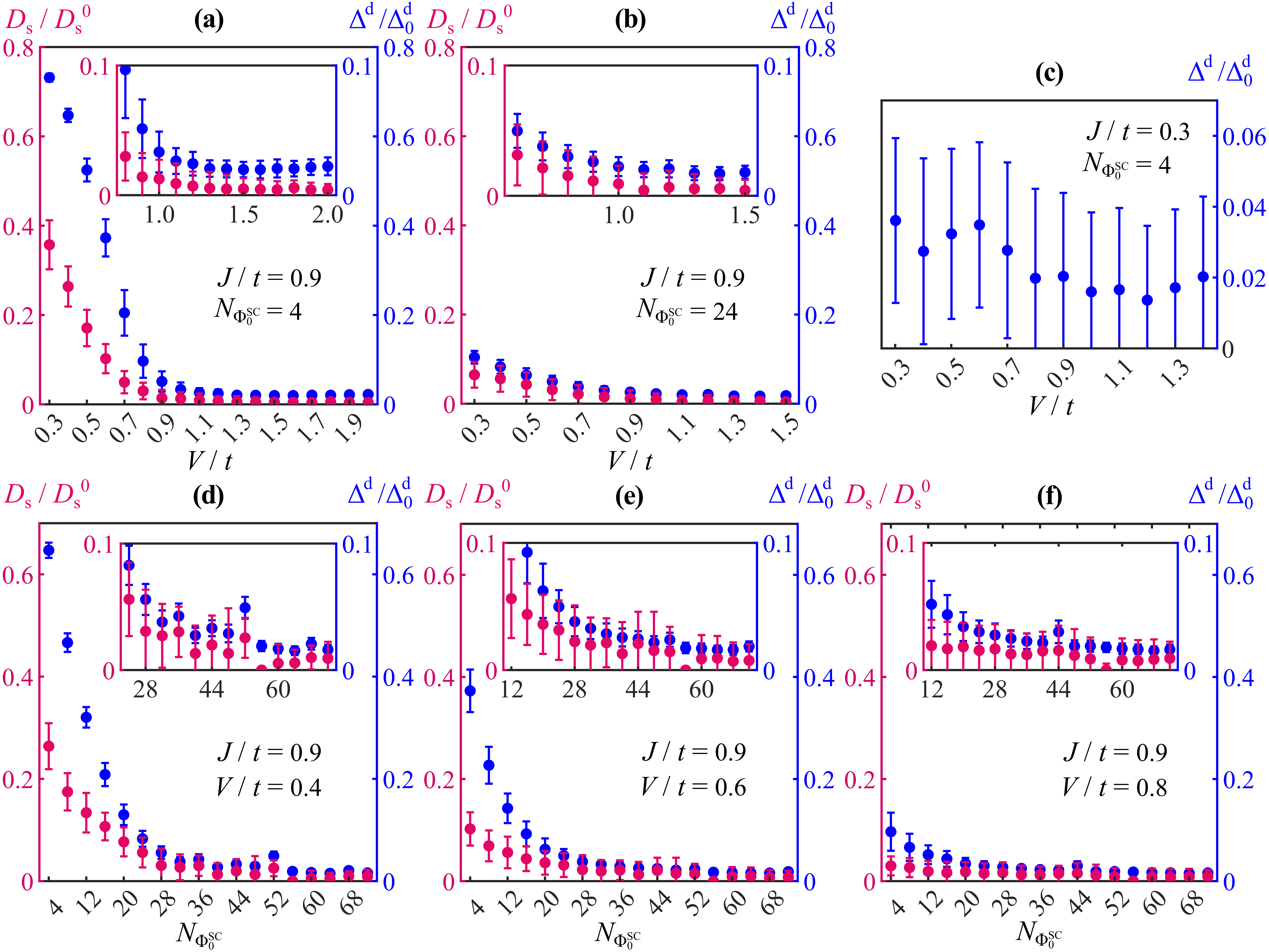}
    \caption{Panel~(a) and (b): The superconducting pairing $\Delta^{\text{d}}$ and superfluid stiffness $D_{\text{s}}$ under increasing disorder for a system penetrated by $4$ and $24$ vortices, respectively, for pairing potential $J/t = 0.9$. 
    Panel~(c): The superconducting pairing for $4$ vortices for a weaker pairing potential $J/t = 0.3$.
    Panel~(d), (e), and~(f): The superconducting pairing and superfluid stiffness as a function the number of vortices penetrating the system for disorder strengths $V/t = 0.4$, $V/t = 0.6$, and $V/t = 0.8$ for pairing potential $J/t = 0.9$.
    In all panels, the superconducting pairing and superfluid stiffness are averaged over all lattice sites and impurity configurations, and plotted with respect to their values $\Delta_0^{\text{d}}$ and $D_{\text{s}}^0$ in a clean system with zero external magnetic field. The insets show a zoom in on the data in the main plots. The error bars represent the standard deviation.
    (In this Figure, $N_x = N_y = 28$, and $M_x = M_y = 1$.)}
    \label{fig:03}
\end{figure}

In Fig.~\ref{fig:03}(a), (b) and~(c), we plot the average superconducting pairing and superfluid stiffness as a function of disorder for a system under a constant applied magnetic field.
In panel~(a) where the applied magnetic field is weak, the superconducting pairing remains finite for all impurity strengths and configurations, while the superfluid stiffness goes to zero for an increasing fraction of the impurity configurations as the disorder is increased. For strong disorder, we thus find a regime of finite superconducting pairing beyond the superconducting transition as was predicted in Ref.~\cite{li_npj_21} for zero applied magnetic field. The superconducting pairing survives in islands that grow smaller and fewer in number as the disorder increases. Such islands survives for much higher disorder strengths than what is presented in the figure.
In panel~(b), we consider an applied field that is closer to the critical field of the superconductor. While the superfluid stiffness is still more suppressed than the superconducting pairing, the difference is smaller than for weaker field strengths.
In panel~(c), we show the average superconducting pairing as a function of the disorder strength for a weaker pairing potential. This demonstrates that there is a second transition where the superconducting pairing also goes to zero. However, this transition happens for a higher disorder strength and magnetic field than what is reasonable to consider for the band width and system size in panel~(a) and~(b).

In Fig.~\ref{fig:03}(d), (e) and~(f) we study the average superconducting pairing and superfluid stiffness as a function of the applied magnetic field for different disorder strengths.
Despite the non-monotonous behavior caused by the small system size, we see that the average pairing always remain finite, while some fraction of the impurity configurations result in zero superfluid stiffness for the higher field strengths, similar to the results in panel~(a) and~(b).
These results differs qualitatively for what is expected for a clean system, where we know that the superconducting pairing and superfluid stiffness must go to zero simultaneously at the critical field.
Instead, we find that in disordered systems, the intermediate granular regime appears also beyond the magnetic field driven superconducting transition.  In panel~(f), where the disorder strength is high, the superfluid stiffness starts its transition at lower field strengths than in panel~(d) and~(e). This, together with the results from panel~(a) and~(b), indicates that the intermediate granular regime appears at lower field strengths with increasing disorder.
As shown in Fig.~\ref{fig:02}, the vortices contribute to suppressing superconductivity in the already disordered regions and thus makes the superconductivity granular. Once the system is granular, the magnetic field does not punch additional holes in the superconducting condensate and the pairing decreases very slowly as the field is increased.
It is interesting to note that the separation between the two transitions where the superfluid stiffness and superconducting pairing vanishes, also found for a conventional $s$-wave superconductor in Ref.~\cite{datta_arxiv_21}, persists despite our conservative choice of parameters where the flat bands makes the $d$-wave pairing very sensitive to impurity scattering. 
As a result, the intermediate regime of remnant superconducting pairing in the superconductor--metal transition can be conveniently studied by tuning the external magnetic field, provided that the system is sufficiently disordered.

\section{Concluding remarks}

We have here provided a description of the magnetic field driven superconductor--metal transition in the disordered hole-overdoped cuprates when described solely within the dirty-BCS theory. We find that the CdGM zero-bias peak in the local density of states at the vortex cores vanishes already at moderate disorder, due to the vortices being attracted to the most disordered regions. We also show that there is an intermediate regime with remnant superconducting pairing at the superconductor--metal transition, which can be reached by tuning an external magnetic field. 
It still debated to what extent the more unconventional nature of the cuprates needs to be taken into account in the description of the hole-overdoped regime. 
While we have here studied the low-temperature limit, it is likely that at temperatures closer to the critical temperature, thermal fluctuations could be the dominant cause for the loss of phase coherence of the Cooper pairs. 
Moreover, experiments predict a pseudogap to exist in the antinodal regions of the Fermi surface above the superconducting critical temperature, particularly in underdoped to weakly overdoped samples \cite{fradkin_rmp_15}. It is unclear whether a second pseudogap could enter the density of states also beyond the disorder and field driven superconducting transition in the highly disordered overdoped samples considered here.
Another open question is what the exact nature of the material is at disorder and field strengths where both the superfluid stiffness and superconducting pairing is absent. Experimental studies suggest that the hole-overdoped cuprates are metallic rather than insulating in the normal-state suggesting that the material could be conducting beyond the two transitions \cite{bozovic_jsnm_18}.
Although it far outside the scope of this work to resolve this debate, we find that the superconductor--metal transition in the disordered hole-overdoped cuprates show some interesting features even when described within the BdG framework. 


\section*{Acknowledgments}

This work was supported by the Research Council of Norway through its Centres of Excellence funding scheme, Project No. 262633 "QuSpin".


\appendix

\section{Theoretical framework}
\label{appendix}

\subsection{The full Bogoliubov--de Gennes equations}

In order to diagonalize the Hamiltonian in Eq.~\eqref{eq:Hamiltonian}, we solve the Bogoliubov-de Gennes equations of the system following the approach in Refs.~\cite{zhu_book_16,hermansen_thesis_18}. 
We first define a basis
\begin{align}
    \psi_{\ve{i}}=\big(c_{\ve{i},\uparrow}\:\:c_{\ve{i},\downarrow}\:\:c^{\dagger}_{\ve{i},\uparrow}\:\:c_{\ve{i},\downarrow}^{\dagger}\big)^{\text{T}} ,
\end{align}
and write the Hamiltonian in the form
\begin{align}
    H = H_0 +\frac{1}{2}\sum_{\ve{i},\ve{j}}\psi_{\ve{i}}^{\dagger}H_{\ve{i},\ve{j}}\psi_{\ve{j}}.
\end{align}
Above, $H_0$ is a constant and $H_{\ve{i},\ve{j}}$ is a $4\times4$ matrix. The Hamiltonian can be written in a diagonal form 
\begin{align}
    H = H_0 +\frac{1}{2}\sum_{n}E_n \gamma_n^{\dagger}\gamma_n 
\end{align}
by solving the full BdG equations
\begin{align}
    \sum_{\ve{j}}H_{\ve{i},\ve{j}}\phi_{\ve{j},n} = E_n \phi_{\ve{i},n}.
\end{align}
Here, $E_n$ are the eigenenergies and $\phi_{\ve{i},n}$ the eigenvectors labeled by $n\in[1,4N_x N_y ]$. There are seemingly twice as many fermionic operators $\gamma_n$ compared to our original operators $c_{\ve{i},\sigma}$, which means pairs of the new operators must related to each other. It can be shown that there are two equivalent solutions
\begin{align}
    E_n \:,\:\phi_{\ve{i},n}&=\big(u_{\ve{i},n\uparrow}\:\:u_{\ve{i},n\downarrow}\:\:v_{\ve{i},n\uparrow}\:\:v_{\ve{i},n\downarrow}\big)^{\text{T}}   ,\\
    -E_n \ ,\: \phi_{\ve{i},n}&=\big(v^*_{\ve{i},n\uparrow}\:\:v^*_{\ve{i},n\downarrow}\:\:u^*_{\ve{i},n\uparrow}\:\:u^*_{\ve{i},n\downarrow}\big)^{\text{T}}  .
\end{align}
Since the eigenenergies of these equivalent solutions differ only by a sign, we can write the the Hamiltonian in a diagonal form
\begin{align}
    H=H_0 -\frac{1}{2}\sum_{n\text{ for }E_n >0}E_n +\sum_{n\text{ for }E_n >0}E_n \gamma^{\dagger}_{n}\gamma_n 
\end{align}
including only positive eigenenergies. The old operators are related to the new ones by
\begin{align}
    c_{\ve{i},\sigma}=\sum_{n\text{ for }E_n >0}\big(u_{\ve{i},n,\sigma}\gamma_n +v_{\ve{i},n,\sigma}^* \gamma_n^{\dagger} \big).
\end{align}
Since the operators in the diagonalized Hamiltonian are now independent, the expectation values of the new operators can be evaluated as
\begin{align}
    &\left<\gamma_n^{\dagger}\gamma_m \right>=
    f_{\text{FD}}(E_n )\delta_{n,m},\\
    &\left<\gamma_n^{\dagger}\gamma_m^{\dagger} \right>=\big<\gamma_n \gamma_m \big>=0
\end{align}
for $E_n >0$, where $f_{\text{FD}}(E_{n})$ is the Fermi-Dirac distribution.

\subsection{The reduced Bogoliubov--de Gennes equations}

We can simplify our calculation by realizing that in the absence of spin-orbit coupling and spin-flip scattering, the Hamiltonian matrix contains two independent blocks \cite{zhu_book_16}. It turns out that the two independent sets of BdG equations can be written in exactly the same form and that while one results in positive eigenenergies, the other results in negative eigenenergies. It is therefore far more efficient to solve the reduced BdG equations 
\begin{align}
    \sum_{\ve{j}}
    \begin{pmatrix}
    \epsilon_{\ve{i},\ve{j}} & \Delta_{\ve{i},\ve{j}} \\
    \Delta^*_{\ve{j},\ve{i}} & -\epsilon_{\ve{j},\ve{i}}
    \end{pmatrix}
    \begin{pmatrix}
    u_{\ve{j},n}\\
    v_{\ve{j},n}
    \end{pmatrix}
    =E_n
    \begin{pmatrix}
    u_{\ve{i},n}\\
    v_{\ve{i},n}
    \end{pmatrix}
\end{align}
for all positive and negative eigenenergies labeled by $n\in[1,2N_x N_y ]$. For simplicity of notation, we have defined 
\begin{align}
    \epsilon_{\ve{i},\ve{j}}=-t_{\ve{i},\ve{j}}e^{i\phi_{\ve{i},\ve{j}}}-(\mu-V_{\ve{i}})\delta_{\ve{i},\ve{j}}.
\end{align}
The diagonalized Hamiltonian can then be written in the form
\begin{align}
    H=H_0 -\frac{1}{2}\sum_{n}|E_n | +\sum_{n} |E_n | \gamma^{\dagger}_{n}\gamma_n ,
\end{align}
where the old operators can be written in terms of new operators using the relations
\begin{align}
    c_{\ve{i},\uparrow}&=\sum_{n \text{ for }E_n >0}u_{\ve{i},n}\gamma_n +\sum_{n \text{ for }E_n <0}u_{\ve{i},n}\gamma_n^{\dagger},\\
    c_{\ve{i},\downarrow}&=\sum_{n \text{ for }E_n >0}v^*_{\ve{i},n}\gamma_n^{\dagger} +\sum_{n \text{ for }E_n <0}v^*_{\ve{i},n}\gamma_n .
\end{align}
The expectation values of the new operators are given by
\begin{align}
    &\left<\gamma_n^{\dagger}\gamma_m \right>=
    f_{\text{FD}}(|E_n |)\delta_{n,m},\\
    &\left<\gamma_n^{\dagger}\gamma_m^{\dagger} \right>=\big<\gamma_n \gamma_m \big>=0.
\end{align}
The reduced BdG equations in their current form is suitable for studying systems of a finite size. However, when studying vortex formation, it is beneficial to consider a larger systems. Therefore, we next introduce periodic boundary conditions to eliminate edge effects. 

\subsection{Boundary conditions and self-consistent solution}

When applying an external magnetic field perpendicular to the sample, the translational invariance of the lattice is broken by the Peierls phase. However, by introducing magnetic unit cells containing an even number of superconducting flux quanta, we can regain the translational invariance of the lattice under translation between equivalent sites in different magnetic unit cells \cite{zhu_book_16,han_prb_00,schmid_njp_10}.  This allows us to use periodic boundary conditions, and we can consider smaller lattice sizes without the disturbance of edge effects.

We consider $M_x \times M_y$ magnetic unit cells of size $N_x \times N_y$. A translation between magnetic unit cells is described by a vector $\ve{R}_{l_x ,l_y }=(l_x N_x a,l_y N_y a,0)$, where $l_{x(y)}\in[0,M_{x(y)}-1]$ and $a$ is the lattice constant. By applying periodic boundary conditions through the magnetic Bloch theorem \cite{brown_pr_64}, our eigenvectors and eigenenergies acquire an index 
\begin{align}
\ve{k}=\frac{2\pi l_x }{M_x N_x a} \ve{x}+\frac{2\pi l_y }{M_y N_y a} \ve{y}.
\end{align}
This allows us to solve the BdG equations for a system size of $N_x \times N_y$ for $M_x M_y$ values of $\ve{k}$, rather than for a system of size $N_x M_x \times N_y M_y $. 
We choose to absorb a $\ve{k}$ dependent phase factor into the eigenvector so that 
\begin{align}
    \begin{pmatrix}
    u_{\ve{i},n,\ve{k}}\\
    v_{\ve{i},n,\ve{k}}
    \end{pmatrix}=e^{i\ve{k}\cdot\ve{r}_{\ve{i}}}
    \begin{pmatrix}
    \tilde{u}_{\ve{i},n,\ve{k}}\\
    \tilde{v}_{\ve{i},n,\ve{k}}
    \end{pmatrix}.
\end{align}
The BdG equations now take the form
\begin{align}
    \sum_{\ve{j}}e^{i\ve{k}\cdot(\ve{r}_{\ve{j}}-\ve{r}_{\ve{i}})}
    \begin{pmatrix}
    \epsilon_{\ve{i},\ve{j}} & \Delta_{\ve{i},\ve{j}} \\
    \Delta_{\ve{j},\ve{i}}^* & -\epsilon_{\ve{j},\ve{i}}
    \end{pmatrix}
    \begin{pmatrix}
    \tilde{u}_{\ve{j},n,\ve{k}} \\
    \tilde{v}_{\ve{j},n,\ve{k}}
    \end{pmatrix}
    =E_{n,\ve{k}}
    \begin{pmatrix}
    \tilde{u}_{\ve{i},n,\ve{k}} \\
    \tilde{v}_{\ve{i},n,\ve{k}}
    \end{pmatrix}.
    \label{eq:BdG_new}
\end{align}
These are solved together with the self-consistency equation for the superconducting pairing correlations
\begin{align}
    \Delta_{\ve{i},\ve{j}}=\frac{U}{M_x M_y}\sum_{n,\ve{k}}e^{i\ve{k}\cdot(\ve{r}_{\ve{i}}-\ve{r}_{\ve{j}})}\tilde{u}_{\ve{i},n,\ve{k}}(\tilde{v}_{\ve{j},n,\ve{k}})^* [1-f_{\text{FD}}(E_{n,\ve{k}})].
    \label{eq:self_consistency_01}
\end{align}

Inside a magnetic unit cell, the only phase factors we need to consider are the Peierls phases associated with electron hopping. The nonzero Peierls phases are $\phi_{\ve{i}\pm\ve{y},\ve{i}}=\mp \pi\phi i_x $ for nearest neighbor hopping, $\phi_{\ve{i}\pm\ve{x}\pm'\ve{y},\ve{i}}=\mp' \pi\phi(i_x \pm1/2)$ for next nearest neighbor hopping, and $\phi_{\ve{i}\pm\ve{y},\ve{i}}=\mp 2\pi\phi i_x $ for third nearest neighbor hopping. These depend on the magnetic field through $\phi=N_{\Phi_0^{\text{SC}}}/N_x N_y  = Ba^2 / \Phi_0^{\text{SC}}$.
When site $\ve{i}$ and $\ve{j}$ in Eq.~\eqref{eq:BdG_new} and~\eqref{eq:self_consistency_01} lies in different magnetic unit cells, we need to apply the translation $\ve{R}_{l_x ,l_y }$ to one of the eigenvectors so that all eigenvalues lie in the same magnetic unit cell. Upon such a translation, the eigenvalues pick up an additional phase through the boundary condition
\begin{align}
    \begin{pmatrix}
    \tilde{u}_{\ve{i},n,\ve{k}}(\ve{r}_{\ve{i}}+\ve{R}_{l_x ,l_y }) \\
    \tilde{v}_{\ve{i},n,\ve{k}}(\ve{r}_{\ve{i}}+\ve{R}_{l_x ,l_y })
    \end{pmatrix}
    =
    \begin{pmatrix}
    e^{-i\chi(\ve{r}_{\ve{i}},\ve{R}_{l_x ,l_y })/2}\tilde{u}_{\ve{i},n,\ve{k}}(\ve{r}_{\ve{i}}) \\
    e^{+i\chi(\ve{r}_{\ve{i}},\ve{R}_{l_x ,l_y })/2}\tilde{v}_{\ve{i},n,\ve{k}}(\ve{r}_{\ve{i}})
    \end{pmatrix}.
    \label{eq:boundary_condition}
\end{align}
The phase
\begin{align}
    \chi(\ve{r}_{\ve{i}},\ve{R}_{l_x ,l_y})=\frac{2\pi}{\Phi_0^{\text{SC}}}\ve{A}(\ve{R}_{l_x ,l_y })\cdot\ve{r}_{\ve{i}}=2\pi\phi l_x N_x i_y .
\end{align}
is the total phase picked up by the superconducting pairing through the translation
\begin{align}
    \Delta^{\text{d}}_{\ve{i}}(\ve{r}_{\ve{i}} - \ve{R}_{l_x ,l_y }) = \Delta^{\text{d}}_{\ve{i}}(\ve{r}_{\ve{i}} )e^{i\chi(\ve{r}_{\ve{i}},\ve{R}_{l_x ,l_y })}.
\end{align}
By solving the reduced BdG equation in Eq.~\eqref{eq:BdG_new} together with the self-consistency equation in Eq.~\eqref{eq:self_consistency_01} and the boundary condition in Eq.~\eqref{eq:boundary_condition}, we obtain eigenenergies and eigenvalues that we can use to calculate physical observables. 

\subsection{Physical observables}

We here give the expressions for the physical observables in terms of the eigenenergies and eigenvalues. The hole concentration is given by
\begin{align}
    x = \frac{1}{N_x N_y M_x M_y}\sum_{\ve{i},n,\ve{k}}\big\{1-|\tilde{u}_{\ve{i},n,\ve{k}}|^2 f_{\text{FD}}(E_{n,\ve{k}})-|\tilde{v}_{\ve{i},n,\ve{k}}|^2 \big[1-f_{\text{FD}}(E_{n,\ve{k}})\big]\big\}
\end{align}
and determines the doping level. 
The $d$-wave superconducting pairing is calculated using Eq.~\eqref{eq:dwave} and Eq.~\eqref{eq:self_consistency_01}.
The superfluid stiffness is calculated from the Kubo formula in Eq.~\eqref{eq:Kubo1}, where we insert the expectation value of the kinetic energy associated with the $x$ oriented bonds
\begin{align}
    \left<-K_x \right>=&\frac{1}{N_x N_y M_x M_y}\sum_{\ve{i},\ve{\delta},n,\ve{k}}\delta_x^2 \big(t_{\ve{i}+\ve{\delta},\ve{i}}e^{i\phi_{\ve{i}+\ve{\delta},\ve{i}}}\big\{\tilde{u}_{\ve{i}+\ve{\delta},n,\ve{k}}^* \tilde{u}_{\ve{i},n,\ve{k}}e^{-i\ve{k}\cdot\ve{\delta}}f_{\text{FD}}(E_{n})\notag\\
    &+\tilde{v}_{\ve{i},n,\ve{k}}^* \tilde{v}_{\ve{i}+\ve{\delta},n,\ve{k}}e^{i\ve{k}\cdot\ve{\delta}}\big[1-f_{\text{FD}}(E_{n})\big]\big\} + \text{c.c}\big)
    \label{eq:kinetic}
\end{align}
and the current-current correlation function
\begin{align}
    \Lambda_{xx}(\ve{q},\omega)=\frac{1}{N_x N_y (M_x M_y )^2}\sum_{n,\ve{k},m,\ve{k}'}\frac{f_{\text{FD}}(E_{n,\ve{k}})-f_{\text{FD}}(E_{m,\ve{k}'})}{\omega+i\delta +E_{n,\ve{k}}-E_{m,\ve{k}'}}A_{n,\ve{k},m,\ve{k}'}(-\ve{q})A_{m,\ve{k}',n,\ve{k}}(\ve{q}).
\end{align}
Above, $\ve{\delta}=\{\ve{x},\ve{y},\ve{x}\pm\ve{y},2\ve{x},2\ve{y}\}$ and 
\begin{align}
    A_{m,\ve{k}',n,\ve{k}}(\ve{q})=\sum_{\ve{i},\ve{\delta}}\delta_x t_{\ve{i}+\ve{\delta},\ve{i}}e^{i(\ve{q}-\ve{k}'+\ve{k})\cdot\ve{r}_{\ve{i}}}
    \big[\big(&\tilde{u}_{\ve{i}+\ve{\delta},m,\ve{k}'}^* \tilde{u}_{\ve{i},n,\ve{k}}e^{-i\ve{k}'\cdot\ve{\delta}}
    -\tilde{v}_{\ve{i},m,\ve{k}'}^* \tilde{v}_{\ve{i}+\ve{\delta},n,\ve{k}}e^{i\ve{k}\cdot\ve{\delta}}\big)e^{i\phi_{\ve{i}+\ve{\delta},\ve{i}}} \notag\\
    +\big(&\tilde{v}_{\ve{i}+\ve{\delta},m,\ve{k}'}^* \tilde{v}_{\ve{i},n,\ve{k}}e^{-i\ve{k}'\cdot\ve{\delta}}
    -\tilde{u}_{\ve{i},m,\ve{k}'}^* \tilde{u}_{\ve{i}+\ve{\delta},n,\ve{k}}e^{i\ve{k}\cdot\ve{\delta}}\big)e^{-i\phi_{\ve{i}+\ve{\delta},\ve{i}}}\big].
\end{align}
The bond currents can be obtained by multiplying Eq.~\eqref{eq:kinetic} with $i$ and reversing the sign of the complex conjugate. In Fig.~\ref{fig:02}, we have included bond currents along all bonds by removing the factor $\delta_x^2$.
Finally, the local density of states is given by 
\begin{align}
    N_{\ve{i}} = \frac{1}{M_x M_y}\sum_{n,\ve{k}}\big[|\tilde{u}_{\ve{i},n,\ve{k}}|^2 \delta(\omega-E_{n,\ve{k}})+|\tilde{v}_{\ve{i},n,\ve{k}}|^2 \delta(\omega+E_{n,\ve{k}})\big].
\end{align}
In our numerical calculation, the $\delta$-function is approximated by $\delta(x)=(1/\pi)[\Gamma/(x^2 +\Gamma^2 )]$, where $\Gamma=0.05$.

\vspace{2cm}


\end{document}